\documentclass[aps,prl,superscriptaddress,amsmath,amssymb,aps,twocolumn,]{revtex4-2}
\usepackage{graphicx}
\usepackage{dcolumn}
\usepackage{bm}
\usepackage{latexsym,epsfig}
\usepackage{graphicx}
\usepackage{verbatim}
\usepackage{comment}
\usepackage{amsmath}
\usepackage{amsfonts}
\usepackage{amsthm}
\usepackage{amssymb}
\usepackage{mathrsfs}
\usepackage{stmaryrd}
\usepackage{color}
\usepackage{physics}
\usepackage{epstopdf}
\usepackage{grffile}
\usepackage{bbold}
\usepackage[normalem]{ulem}
\usepackage{mathtools}
\DeclareGraphicsExtensions{.eps}
\newcommand{\beq}{\begin{equation}}
\newcommand{\eeq}{\end{equation}}
\newcommand{\bea}{\begin{eqnarray}}
\newcommand{\eea}{\end{eqnarray}}
\newcommand{\ben}{\begin{eqnarray*}}
\newcommand{\een}{\end{eqnarray*}}
\newcommand{\bfig}{\begin{figure}}
\newcommand{\efig}{\end{figure}}

\newcommand{\ra}{\rangle}

\newcommand{\non}{\nonumber}
\newcommand{\al}{\alpha}

\newcommand{\ga}{\gamma}

\newcommand{\da}{\dagger}

\usepackage{hyperref}
\hypersetup{
 colorlinks=true, 
 urlcolor=blue,
 citecolor=blue,
 linkcolor=blue
}

\begin{document}
\title{Emergence of a molecular quantum liquid in one dimension} 
\date{\today}
\author{Rajashri Parida}
\email{rajashriparida33@gmail.com}
\affiliation{School of Physical Sciences, National Institute of Science Education and Research, Jatni 752050, India}
\affiliation{Homi Bhabha National Institute, Training School Complex, Anushaktinagar, Mumbai 400094, India}

\author{Biswajit Paul}
\email{biswajitt.bp@gmail.com}
\affiliation{School of Physical Sciences, National Institute of Science Education and Research, Jatni 752050, India}
\affiliation{Homi Bhabha National Institute, Training School Complex, Anushaktinagar, Mumbai 400094, India}

\author{Harish S. Adsule}
\affiliation{JILA and Department of Physics, University of Colorado, Boulder, Colorado 80309, USA}

\author{Shovan Dutta}
\email{shovan.dutta@rri.res.in}
\affiliation{Raman Research Institute, Bangalore 560080, India}

\author{Diptiman Sen}
\email{sen.diptiman@gmail.com}
\affiliation{Centre for High Energy Physics, Indian Institute of Science, Bengaluru 560012, India}

\author{Tapan Mishra}
\email{mishratapan@gmail.com}
\affiliation{School of Physical Sciences, National Institute of Science Education and Research, Jatni 752050, India}
\affiliation{Homi Bhabha National Institute, Training School Complex, Anushaktinagar, Mumbai 400094, India}

\author{Adhip Agarwala}
\email{adhip@iitk.ac.in}
\affiliation{Department of Physics, Indian Institute of Technology, Kanpur 208016, India}

\begin{abstract}
 We investigate the fate of a one-dimensional lattice superfluid formed by hard-core bosons, aka `atoms' (alternatively, a free spinless Fermi sea), subjected to nearest-neighbor attractive Hubbard-like interactions only in subgroups of two sites. The system, as expected, stabilizes a fluid of dimerized molecules at large attractive interactions. However, the composite molecules have an effective meek hopping scale and dominant repulsive interactions solely due to virtual quantum fluctuations. Interestingly, at an intermediate attractive potential, the system realizes a 
 phase-separated region where the system is in an absorbing state. We show that this phase-separated region is due to an emergent attractive interaction between the dimers, which leads to a local charge-density wave puddle where particles effectively cluster with local half-filling. Moreover, the molecular superfluid gets spontaneously charge-ordered in the addition of an unpaired atom, reflecting the extreme sensitivity of the system to the existence of lone atoms. Using density-matrix renormalization group studies and effective low-energy Hamiltonians, we isolate the quantum processes to uncover the physics behind molecule formation in a strongly interacting 
 one-dimensional system. 
\end{abstract}
\maketitle

\paragraph*{Introduction.-} 
\label{sec:intro}
Composite particles emerge across myriad phases of quantum matter, be it in fractional quantum Hall physics, superconductivity, or in quantum spin liquids 
\cite{jain2007composite, stormer1999fractional, savary2016quantum, tinkham2004introduction, broholm2020quantum, ZhouRMP2017, varma2020colloquium}. Often, the ultraviolet degrees of freedom combine in non-trivial ways to lead to quasiparticles that are non-local in the original degrees of freedom. While many such phenomena have been discovered and studied in solid state platforms \cite{lu2024fractional, ju2024fractional, chang2023colloquium}, the advent of synthetic platforms such as cold-atoms and photonic systems \cite{anglin2002bose, cirac2003manipulate, bloch_rev_2000, chemical_review,zhang2018topological, halimeh2025cold, noh2016quantum,ion_atom_review, ozawa2019topological, chang2018colloquium,Schafer2020_rev,Baroni2024,3d_photonic_lattice,arbitrary_photonic_lattice,photonic_topo_ins,Hafezi2013,kagome_photonic_lattice,photonic_lieb_lattice} allow us to engineer model Hamiltonians and explore their physics in low dimensions. A particularly interesting direction in this context is the physics of attractive interactions in one dimension. The motivation to explore this direction has been many - formation of (topological) superconductivity in one dimension \cite{RuhmanPRB2017, toposc1, toposc2, toposc3, toposc4, toposc5, toposc6, toposc7, toposc8}, understanding the attractive Hubbard model physics \cite{extended_attrative_hubbard,Kazuhiro_Sano_ebh_model,hubbard_model_book,ebh_model_quarter_filling,wp_dynamics_ebh}, nature of paired superfluidity \cite{ThomasPRA2016}, and uncovering remnants of pseudogap physics important to understand high-$T_c$ physics~\cite{EmeryPRL1990,CaponePRL2002, qu2022spin,1d_cuprate,phonon_mediated_attraction}. It has also been shown that attractive interactions often lead to the physics of coexistence and phase separation \cite{PencPRB1994}, which has connections to the $5/2$-fractional quantum Hall physics~\cite{KanePRX2017}. The realization of recent experiments in synthetic systems with effective pair-wise attraction, induced via Coulomb interactions \cite{hamo2016electron}, have given further impetus to these ideas.

A natural question which arises is - how do stable molecules arise from single particles in presence of short-ranged attractive interactions? Moreover, what is the effective interaction between such molecules, and what phases can they realize? While such `molecular phases', arising solely from attractive interactions, have been little explored in one dimension; in an important work, Gotta~et.~al.~showed that fermions hopping in one dimension along with competing pair-hopping can lead to the Fermi sea transiting to a paired state. This paired state is a coexistent phase which can be described within a two-fluid picture with central charge $c=2$ \cite{GottaPRL2021, GottaPRB2021}. In this work, we will show that attractive interactions, when only acting at subsets of pairs of sites in one dimension, can also lead to the formation of a strongly interacting dimer superfluid state which has emergent both repulsive and attractive interactions, albeit due to distinct origins. Furthermore, we will show that the molecular superfluid emerges from a local charge-density wave, where particles cluster in a finite region for moderate interaction strengths. The system has an intriguing odd-even effect where the addition of a particle destabilizes the dimer-superfluid to a charge-density wave puddle. We also discuss how such a Hamiltonian can be realized using ultracold atoms in an optical lattices.

\paragraph*{Model.-} 
\label{sec:model} We consider the following Hamiltonian in one dimension,
\begin{align}
H &= -t ~\sum_{i=0}^{L-1} \Big( b_i^\dagger b_{i+1} +\text{H.c.} \Big)- U\sum_{i, i \in \text{even}} n_i n_{i+1}, \label{eq:Ham}
\end{align}
where $b^\dagger_i, b_i$ are hard-core bosonic creation and annihilation operators with the number operator $n_i = b^\dagger_i b_i$ and $b^2_i = (b^\dagger_i)^2 =0$. $U$ is the inter-site attractive potential (assumed $>0$) which acts only on alternate pairs of sites. Thus, the interaction term respects a lower translational symmetry via displacements by two lattice constants, unlike the first term. Our interest will be to understand the fate of this model as $U$ is increased. Under a Jordan-Wigner transformation, the model will behave like a model of spinless fermions hopping with nearest-neighbor interactions. The one-dimensional Fermi sea is expected to be stable for weak attractive interactions, and within a Hartree-Fock approximation, the ground state energy density of the system is expected to go as \cite{si}
\beq
\epsilon_{\text{FS}} = -\frac{2 t}{\pi} \sin (\rho \pi) - \frac{1}{2} U \Big[\rho^2 - \Big(\frac{\sin(\rho \pi)}{\pi } \Big)^2\Big]. \label{eq:lowUeq}
\eeq

\paragraph*{Large $U$ limit.-} 
\label{sec:largeU} At $U \rightarrow \infty$ it is expected that pairs of particles will reside at pairs of composite sites. For instance, the two sites $(i=2p, i+1=2p+1)$ form the $p^{\text{th}}$ composite site. Thus, the $L$ sites of the original lattice leads to $L_p = L/2$ composite sites, and the number of original particles $N \rightarrow N_p = \frac{N}{2}$ dimers. The many-body degenerate space of $\binom{L_p}{N_p}$ with energy $-N_p U$ however, gets mixed up due to the perturbative hopping of dimers. The dimer hopping term is found to be $\tilde{t} \simeq \frac{2t^4}{U^3}$ (see SM).  Four hoppings of the original particles  are required to move a dimer from a composite site $p \rightarrow p+1$. Associating a dimer creation operator $d^\dagger_p$ on composite site $p$, the dimer number operator is given by 
\beq n^d_p = d^\dagger_p d_p \equiv n_{2p}n_{2p+1}. \eeq
Interestingly the dimers repel each other strongly, since when a dimer is next to an empty composite site, individual particle can hop virtually, reducing the energy as $-\frac{t^2}{U}$. Thus the full dimer Hamiltonian at 
large-$U$ is 
\beq
H_{\text{eff}} = \sum_{p=0}^{L_p-1} [- \Big( \tilde{t} d^\dagger_{p} d_{p+1} + \text{H.c.} \Big) + V_{\text{eff}} n^d_{p} n^d_{p+1} - \mu_{\text{eff}} n^d_p],
\label{eq:largeUlim}
\eeq
where $V_{\text{eff}} = 2t^2/U$ and $\mu_{\text{eff}}= U + \frac{2t^2}{U}$, thus realizing a $t-V$ model. Given that this is a strongly repulsive hard-core bosonic system in one-dimension, it is expected to stabilize a dimer superfluid (DSF) state away from half-filling.

\begin{figure}
\centering
\includegraphics[width=1.\linewidth]{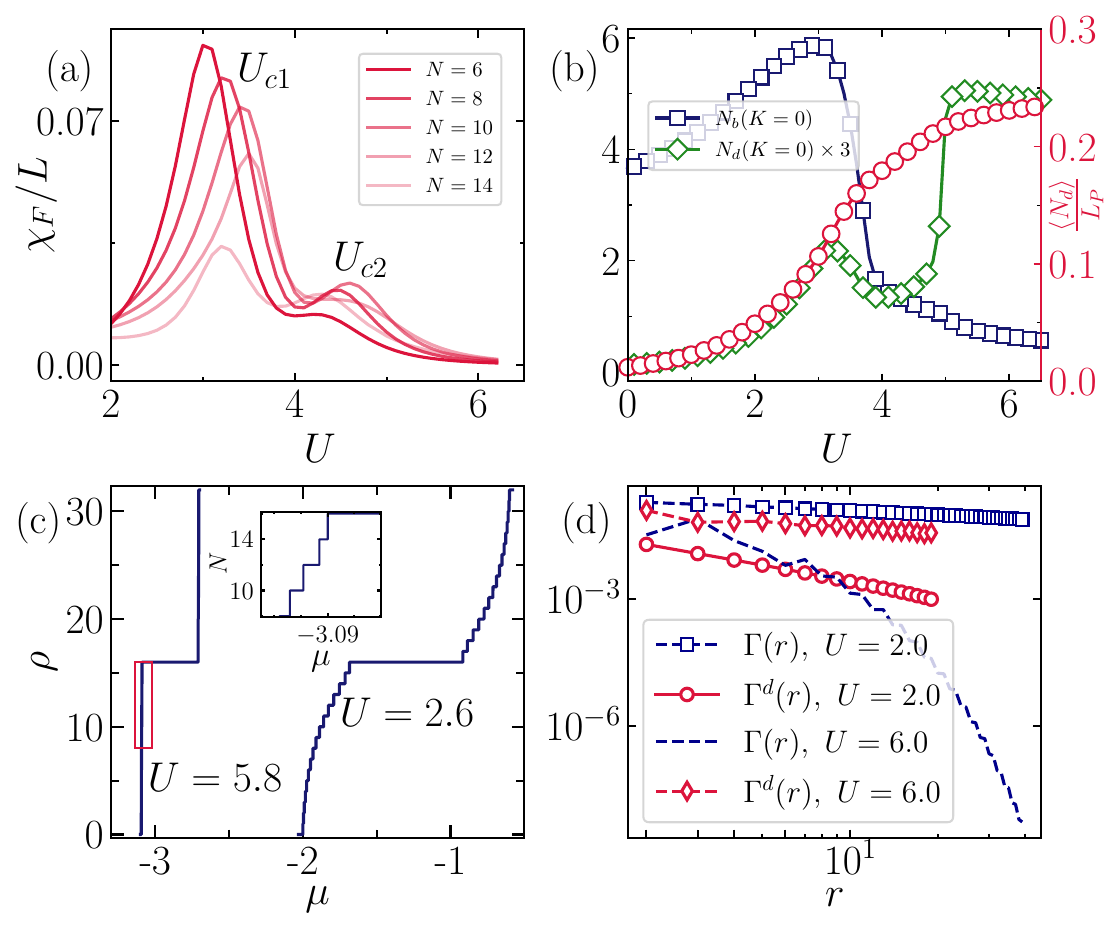}
\caption{{\bf Formation of molecular superfluid:} (a) Fidelity susceptibility $\chi_F$ as a function of $U$ for $L=32$ under OBC at different particle fillings ($N$ particles). (b) Dimer density $\langle N_d \rangle/L_p$ (red circles), SF momentum distribution $\langle N_b(K=0)\rangle$ (blue squares), and DSF momentum distribution $\langle N_d(K=0)\rangle$ (green diamonds) as a function of $U$ for a system of size $L=80$ at $\rho=1/4$ filling. $\langle N_d(K=0)\rangle$ is multiplied by $3$ for better visibility. (c) Particle number as a function of chemical potential $\mu$ for $U=2.6$ and $U=5.8$, respectively, for $L=32$ under PBC. The region indicated by the red rectangle is enlarged in the inset. (d) Single-particle correlation function $\Gamma(r)$ and pair correlation function $\Gamma^{d}(r)$ on a log-log scale at $\rho=1/4$ for $U=2.0$ and $U=6.0$, for $L=80$ under OBC.}
\label{fig:figure_2}
\end{figure}
\begin{figure*}
\centering
\includegraphics[width=1\linewidth]{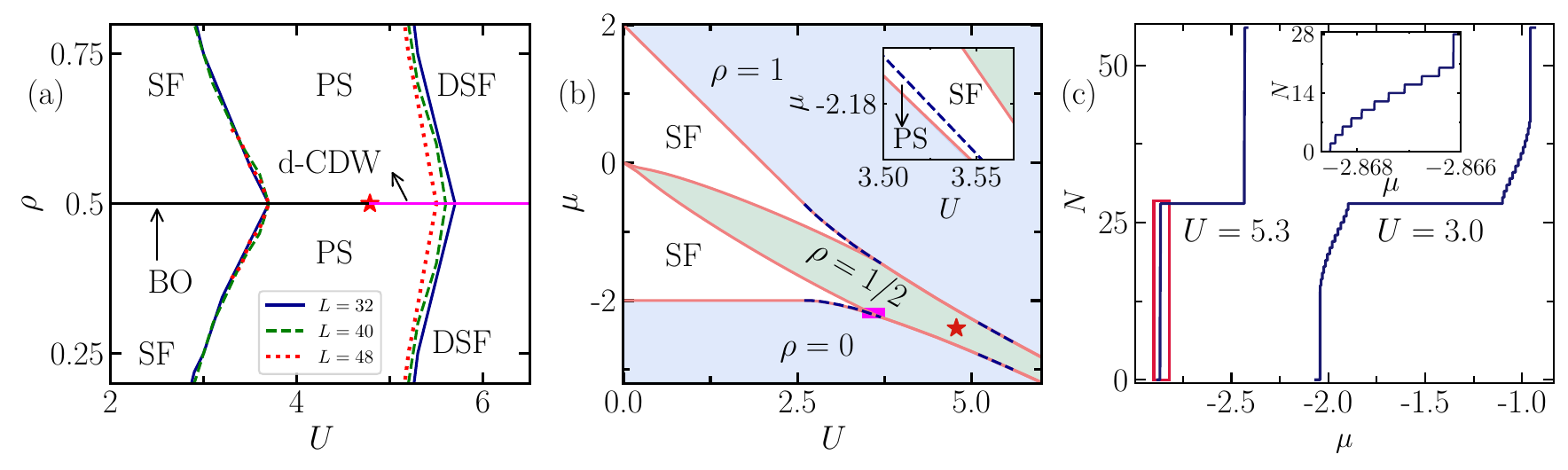}
\caption{ {\bf Phase diagram:} (a) Phase diagram in the
$\rho-U$ plane for different system sizes under PBC. The horizontal solid line at $\rho=0.5$ indicates the gapped phases. (b) Phase diagram of the model in the $U$–$\mu$ plane with PBC. Shaded regions are the gapped phases, and the white regions are gapless phases. The zoomed-in portion of the magenta rectangular region is shown in the inset. The red star in (a) and (b) marks the transition point from the BO to the d-CDW phase at half-filling. (c) shows the particle number as a function of $\mu$ for $U=3.0$ and $U=5.3$, respectively, for $L=56$ under PBC. The zoomed-in version of the red rectangular region is shown in the inset.} \label{fig:figure_3}
\end{figure*}
\paragraph*{Dimer formation.-} 
\label{sec:dimerf}

Having looked at both the small-$U$ and large-$U$ limits, where at partial filling we expect stable superfluid (SF) phases of individual particles and composite particles, 
we now numerically investigate the different phases and phase transitions. The ground state energy of the model (see Eq.~\eqref{eq:Ham}) and its comparison of the effective model at low $U$ (see Eq.~\eqref{eq:lowUeq}) and ground state energy in the large-$U$ limit (see Eq.~\eqref{eq:largeUlim}) shows that the limits are effectively captured (see SM). However, an investigation of the fidelity susceptibility of the system doesn't show a single transition between the two phases; we see in Fig.~\ref{fig:figure_2}(a) that the ground state fidelity susceptibility $\chi_F$ surprisingly shows two peaks, which are thermodynamically stable (see SM). As expected, with increasing $U$ the total number of dimers as given by 
\beq N_d ~=~ \sum_{p=0}^{L_p-1} n^d_p ~\equiv~ \sum_{p=0}^{L_p-1} n_{2p} n_{2p+1} \eeq
shows that the dimer density $\langle N_d \rangle/L_p \rightarrow \rho$ as $U \rightarrow \infty$ (See  Fig.~\ref{fig:figure_2}(b)). Given that $U\rightarrow 0$ and $U \rightarrow \infty$ correspond to superfluids of particles and dimers, respectively, the same features can be captured by the momentum $K=0$ occupancy of the respective degrees of freedom. For the $b$ bosons, defining
\beq
N_b(K) = \frac{1}{L} \Big(\sum_{i=0}^{L-1} \sum_r \langle b^\dagger_i b_{i+r} \rangle \exp(-i K r) \Big), \eeq
and correspondingly for the $d$ dimers,
\beq
N_d(K) = \frac{1}{L_p} \Big(\sum_{p=0}^{L_p-1} \sum_r \langle d^\dagger_p d_{p+r} \rangle \exp(-i K r) \Big),
\eeq
we plot $N_b(K=0)$ and $N_d(K=0)$. We find that while for small $U$, $N_b(K=0) \gg N_d(K=0)$, for large $U$, $N_d(K=0) \gg N_b(K=0)$ (see Fig.~\ref{fig:figure_2}(b)). 
Given that the susceptibility shows two peaks, we have two critical values of $U$, given by $U_{c1}$ and $U_{c2}$. To further confirm the nature of the small-$U$ and large-$U$ phases, we study the system in the grand-canonical ensemble and study the number of particles that get stabilized as the chemical potential $\mu$ is tuned. Interestingly, the nature of excitations can be clearly demarcated for both $U<U_{c1}$ and $U>U_{c2}$. In Fig.~\ref{fig:figure_2}(c) we find that in the former regime, minute changes in $\mu$ change the particle number in units of one ($U=2.6$), while for $U>U_{c2}$ the changes are in units of two ($U=5.8$), signaling that only dimers form the low-energy excitations. We also study the single-particle correlator 
\beq \Gamma (r) = \langle b^\dagger_i b_{i+r} \rangle,
\eeq
and the dimer-dimer correlator 
\beq
\Gamma^d(r) = \langle (b^\dagger_{2i} b^\dagger_{2i+1})( b_{2i+r}b_{2i+1+r} )\rangle. \eeq
Interestingly, we find that for $U<U_{c1}$, $\Gamma(r)$ falls as a power law but for $U>U_{c2}$, $\Gamma(r)$ falls exponentially, signaling a single-particle gap. For $U>U_{c2}$, $\Gamma^d(r)$ falls as a power law showing that coherence 
has been developed between the dimers.
This justifies the use of the term SF for $U<U_{c1}$ and DSF for $U>U_{c2}$. These exact correlator expressions can be calculated in a system with open boundary conditions (see SM for expressions). 

We can calculate the value of $U$ where the behavior of particles changes from a regime in which the particles move independently of each other to a regime where dimers emerge. Looking at the $\rho \rightarrow 0 $ limit and comparing the energies of the atomic superfluid and formation of dimers, shows that $U_{c1} \approx 3.1$.  An explicit two-particle calculation is consistent with this result, and numerical evidence shows that at around $U\sim 2.5$ the susceptibility peak is concomitant with the
emergence of two-particle bound states (see SM for more details) \cite{si}.

\paragraph*{Phase Diagram.-} Having established the SF and the DSF phases, we now investigate the full phase diagram. The phase diagram is shown in Fig.~\ref{fig:figure_3}(a). At half-filling ($\rho = \frac{1}{2}$), the state has a single peak, where it was recently shown \cite{parida2025} that as a function of $U$, the system undergoes a transition from the bond-ordered (BO) phase to a dimer charge-density wave 
(d-CDW) order (see SM). This bond-order develops due to a formation of bonding orbitals where a particle is shared within pairs of consecutive atomic sites. It is important to note that the emergence of gapped phases at half-filling, while expected from the effective dimer Hamiltonian (see Eq.~\eqref{eq:largeUlim}), is unexpected from a simple mean-field theory for the original Hamiltonian (see Eq.~\eqref{eq:Ham}, see SM for details on mean-field). Both the gapped phases are expected to be stable from the attractive bond-ordered phase at small $U$ and the repulsive $t-V$ model at large $U$ as derived from the effective model in Eq.~\eqref{eq:largeUlim}. 

We track how the particle density $(\rho)$ jumps as the chemical potential $\mu$ is tuned in grand-canonical ensemble. One finds that at values of $U$ such as $U\sim 3.0$ and $U \sim 5.3$, the density shows finite jumps before approaching either the SF or the DSF phase (see Fig.~\ref{fig:figure_3}(c)). These finite jumps indicate that the system is in an absorbing state where the system can continue to absorb particles until it reaches a compressible state (SF/DSF) or an incompressible state (BO or d-CDW). We will refer to this intermediate $U$ region as the PS (phase-separated) region, consistent with similar regions reported in other strongly interacting systems where attractive and repulsive potentials compete \cite{qu2022spin}. However, this intermediate-$U$ region is highly non-trivial as we will discuss below. It is interesting to note that the whole of the PS and the DSF regions are rather hard to detect in the $\mu-U$ phase plane [Fig.~\ref{fig:figure_3}(b)], since the hopping scale is perturbatively small. Thus, tiny changes in $\mu$ seem to switch the large-$U$ system between regions with $\rho=0, ~\frac{1}{2}$ and $1$ only. However, careful investigations at fixed $\rho$ clearly show the existence of compressible phases with small widths. The perturbative scales also make DMRG calculations tricky, since the small hopping scales lead to even smaller Kubo gaps, which compete with machine precision if large system sizes are chosen. Therefore, most of the DMRG results we show are limited to system sizes ($\lesssim 100$) where the ground states and excitations are clearly discernible and well-converged. 

\begin{figure}
\centering
\includegraphics[width=1.\linewidth]{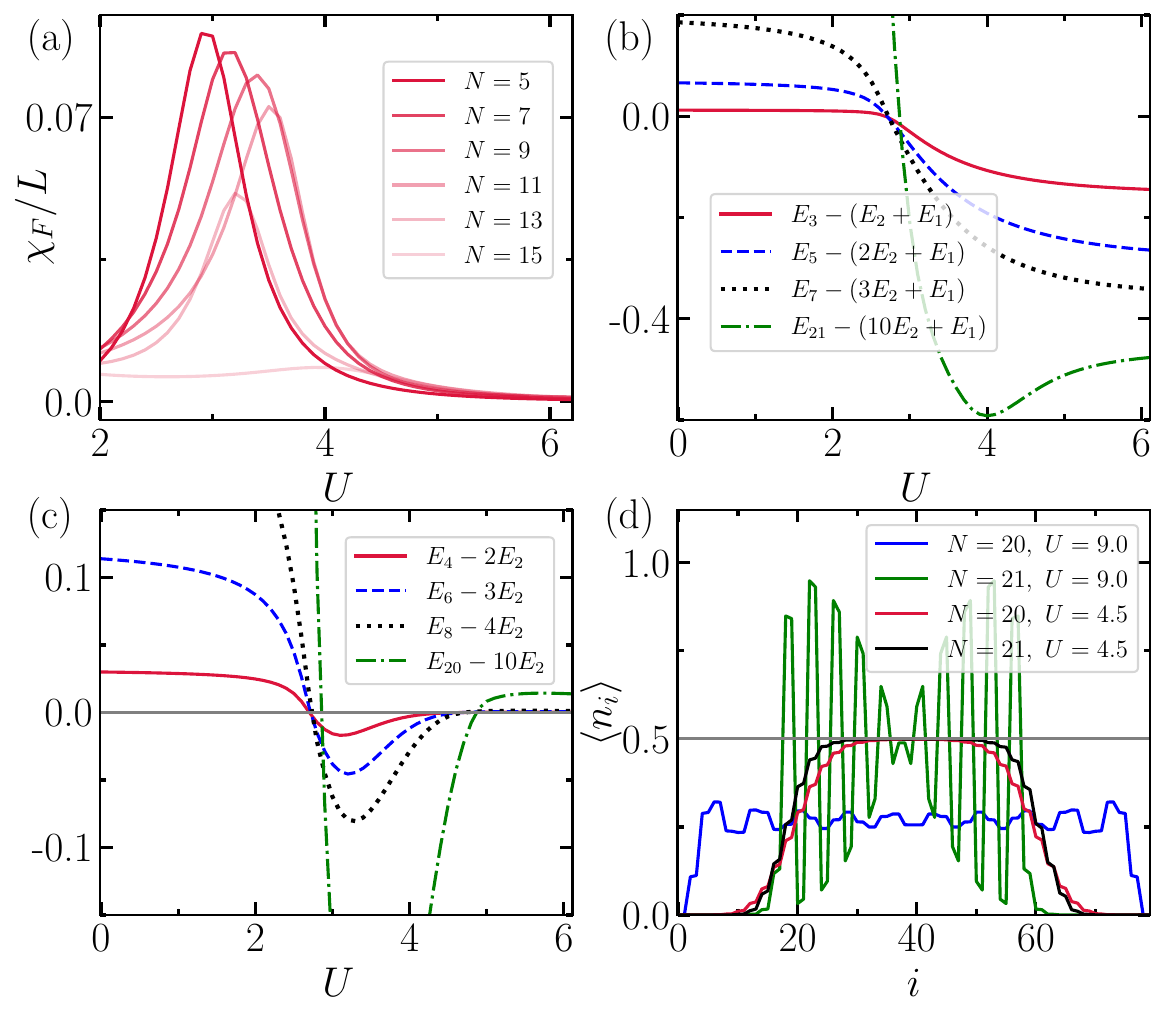}
\caption{{\bf Intermediate-$U$ range:} Fidelity susceptibility, $\chi_F/L$, as a function of $U$ for (a) odd particle filling shown for a system of size $L=32$ calculated under OBC. (b) The ground state energy for odd particle sectors ($E_{2M+1}$) is compared against the ground state energy of $M$ dimers and one lone particle for a system of size $L=80$. (c) The ground state energy for even particle sectors ($E_{2M}$) is compared against the ground state energy of $M$ dimers for a system of size $L=80$. (d) shows the on-site particle occupation for two different values of $U$, when both even and odd numbers of particles are present for a system of size $L=80$.}
\label{fig:figure_4}
\end{figure}

\paragraph*{Intermediate-$U$ range.-} We now discuss the nature of the system in the intermediate-$U$ range, with $U_{c1}<U<U_{c2}$, which resides between the two susceptibility peaks (see~Fig.~\ref{fig:figure_2}(a)). Here, we also report an interesting odd-even effect where we add one extra particle to a commensurate system. In Fig.~\ref{fig:figure_4}(a), one finds that the second susceptibility vanishes as soon as an odd number of particles is added. It seems rather surprising that the addition of just one particle has such a drastic effect on the system. This also sheds light on the nature of the intermediate phase where we find that the system has a propensity to clustering. 

For instance, at $U \rightarrow \infty$, for a 
three-particle system, it might  naively appear that two particles would form a stable dimer while an unpaired particle would hop independently leading to a three-particle energy $E_3= E_2 +E_1 \sim -U -2t$
for $U \gg t$. Here, $E_2$ and $E_1$ are the ground state energies of just two and one particle in the system, respectively. However, an exact calculation (see SM) shows that the three particles prefer to cluster with an energy $E_3 = - U - 2.148 t$ 
for $U \gg t$, thus showing that clustering is favored at large $U$. An explicit calculation of $E_3 - (E_2 +E_1)$ shows that the value goes negative for $U \gtrsim 3$ [Fig.~\ref{fig:figure_4}(b)], showing that the system favors the addition of particles rather than for $U<U_{c1}$ where addition of particles is expensive given the Pauli pressure. In fact this trend continues where for $2M+1$ particles the energy $E_{2M+1} < ME_2 + E_1$ for $U>U_{c1}$ reflecting that the system at large $U$ is not like a set of decoherent dimers and a lone particle, but rather is a strongly correlated state of all particles (see Fig.~\ref{fig:figure_4}(b)). 

The same analysis for an even number of particles illustrates that $E_{2M} - ME_2$ becomes $<0$ only between $U_{c1}$ and $U_{c2}$ (see Fig.~\ref{fig:figure_4}(c)), suggesting that the propensity to clustering now happens only at intermediate $U$. Both for $U<U_{c1}$ and $U>U_{c2}$, this becomes positive as expected, suggesting that a fermionic pressure gets developed in both types of superfluids. However, for $U>U_{c2}$ this value is very close to zero since the quantum pressure is in terms of a perturbative hopping scale $\sim \frac{t^4}{U^3}$. In order to further test any features of clustering, we plot the particle density for both even and odd numbers of particles at $U \gg U_{c2}$ $(U=9.0)$. For an even number of particles, one finds a fairly uniform density, while for an odd number, the particles cluster at the center  (see Fig.~\ref{fig:figure_4}(d)) with prominent density fluctuations. This is in sharp contrast to $U = 4.5$, where for both even and odd numbers of particles, they reside at the center with an effective local density of $\frac{1}{2}$. Note that this local density is different from the actual density $\rho$, which is stabilized both for $U >U_{c2}$ and for $U<U_{c1}$.

\begin{figure}
\centering
\includegraphics[width=1.\linewidth]{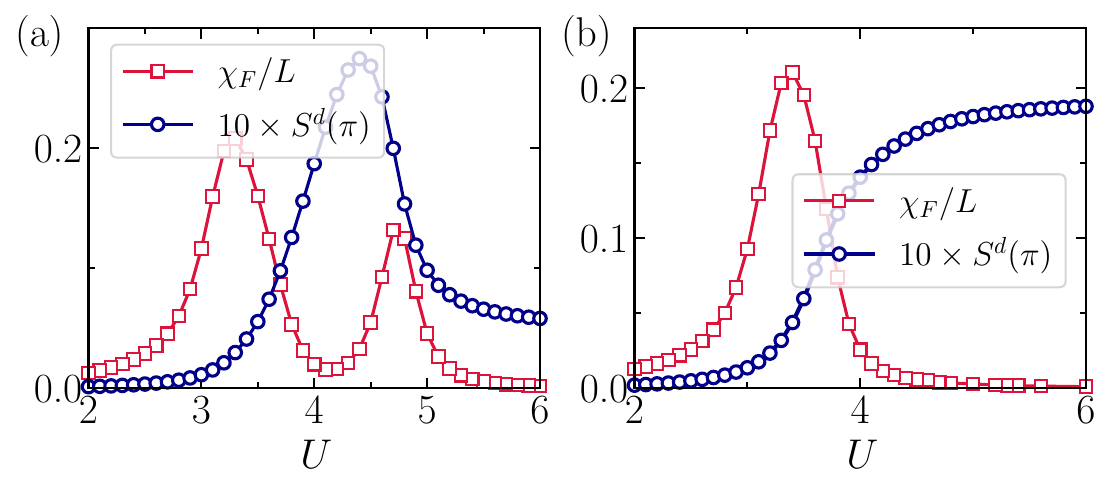}
\caption{ {\bf Local charge order:} $\chi_F/L$ (red squares) and the CDW structure factor $S^d(\pi)$ (blue circles) as functions of $U$, (a) at quarter filling, and (b) for one particle more than quarter filling with $L=48$. All results are obtained with OBC.}
\label{fig:figure_5}
\end{figure}

\paragraph*{Local charge order.-} In order to further find the type of order which gets stabilized in the intermediate $U$ region
where one finds charge clustering but with local density $\rho \sim \frac{1}{2}$, we calculate the structure factor of the dimer-dimer correlator 
\beq
S^d(k) = \frac{1}{L_p^2}\sum_{r,p=1}^{L_p} 
 e^{i k r} \left(\langle n^d_p n^d_{p+r}\rangle
- \langle n^d_p \rangle \langle n^d_{p+r} \rangle\right).
\eeq
When evaluated at intermediate values of $U$, one finds that $S^d(k)$ peaks at $k=\pi$, suggesting that the system develops a charge-density wave (d-CDW) of dimers (see SM).  We find that the strength of $S^d(k=\pi)$ peaks precisely between the two susceptibility peaks of $\chi_F$ (see Fig.~\ref{fig:figure_5}(a)) for an even number of particles showing that the intermediate-$U$ phase is in fact a local CDW puddle of dimers with average density of $\rho=\frac{1}{2}$ lying between the two superfluids. 
This peak behavior disappears for $U>U_{c2}$ (see SM). The same quantity when investigated for an odd number of particles clearly shows that $S^d(\pi)$ is finite at $U_{c1}$ and remains so up to $U \rightarrow \infty$. The rise of $S^d(\pi)$ is concomitant with the susceptibility peak. Thus, it is clear that virtual quantum fluctuations stabilize a local CDW puddle even as a superfluid of monomers transit into a molecular superfluid at large $U$. 

\paragraph*{Outlook.-} The formation of composite particles, as in molecules out of atoms, are often due to Coulombic forces which arise due to charge clouds. In this work, we show that attractive interactions which act only on subgroups of two sites can engineer such composite particles in a one-dimensional system. A natural platform to realize this system would be in an optical lattice with two-component bosons. Our model is equivalent to a XXZ spin-1/2 chain which is realized in the limit of strong on-site interactions at unit filling \cite{Kuklov2003, Jepsen2020}. The staggered $J_z$ coupling can be engineered by using a spin-dependent optical superlattice \cite{altman2003phase, Yang2017}, as detailed in the SM. In particular, this requires the inter- and intra-species interactions ($\equiv U_{\sigma,\sigma'}$) to satisfy $0 < U_{\uparrow\uparrow} < 2 U_{\uparrow\downarrow}$ and $U_{\downarrow\downarrow} < 0$ or $U_{\downarrow\downarrow} > 2 U_{\uparrow\downarrow}$, which is possible with ${}^7$Li atoms~\cite{AmatoGrill2019}.

The composite particles realized via this protocol have both mutual repulsion and attractive interactions at intermediate values of $U$ arising due to quantum fluctuations. 
While the short-range repulsion is due to second-order fluctuations of a single particle, the attractive interactions lead to local charge-density wave order within the composite sites. We find an interesting even-odd effect, where the local CDW clustering remains stable to arbitrarily large $U$ for an odd number of particles. In this work we isolate the
quantum processes which lead to composite particles from constituents. Clearly such processes are difficult to study in any effective (bosonized) theories of the $U \sim 0$ and $U \rightarrow \infty$ limits, showing the need to develop a framework to describe such emergence of quantum molecular liquids in one dimension. Extending this to cases when composite sites are made of odd number of sites and physics of corresponding fermionic phases will be an interesting imminent future study.

\paragraph*{Acknowledgments.-} We acknowledge fruitful discussions with 
Abhishodh Prakash, Ajit C. Balram, Krishendu Sengupta and Arnab Sen. A.A. also acknowledges engaging conversations with Aabhaas V. Mallik, which led to the primary motivations for this work. A.A. acknowledges support from IITK Grants (IITK/PHY/2022010) and (IITK/PHY/2022011).
D.S. thanks SERB, India for funding through Project No. JBR/2020/000043. T.M. acknowledges support from Science and Engineering Research Board (SERB), Govt. of India, through project No. MTR/2022/000382 and STR/2022/000023. This research was supported in part by the International Centre for Theoretical Sciences (ICTS) for participating in the program - Generalised symmetries and anomalies in quantum phases of matter 2026 (code: ICTS/ GSYQM2026/01).
\bibliography{ref}

\newpage
\clearpage

\appendix

\onecolumngrid
\begin{center}
\bf{Supplemental Material for ``Emergence of a molecular quantum liquid in one dimension"}
\end{center}
\twocolumngrid

In the following, we present additional calculations and numerical details supporting the results presented in the main text.

\subsection{Small-$U$ limit}

The energy of the SF at small $U$ can be estimated by first
performing a Jordan-Wigner (JW) transformation to spinless fermions, where 
\begin{align}
H &= -t\sum_i \Big( b_i^\dagger b_{i+1} +\text{H.c.} \Big)- U\sum_{i \in \text{even}} n_i n_{i+1} 
 \\ &\xrightarrow{JW} -t\sum_i \Big( c_i^\dagger c_{i+1} +\text{H.c.} \Big)- U\sum_{i \in \text{even}} n_i n_{i+1}. 
\end{align}
under a mean-field approximation with just the Hartree-Fock terms, we obtain
\beq
H \xrightarrow{\text{Hartree MF}} -t\sum_i \Big( c_i^\dagger c_{i+1} +\text{H.c.} \Big)- U\sum_{i \in \text{even}} \langle n_i n_{i+1}\rangle. \eeq
Assuming a Fermi sea with a uniform density
$\rho$ we know that 
\beq \langle n_i n_{i+r} \rangle = \rho^2 - \Big(\frac{\sin(\rho \pi r)}{\pi r}\Big)^2. \eeq
This leads to the following estimate of the energy density
\beq
\epsilon_{\text{FS}} = -\frac{2 t}{\pi} \sin (\rho \pi) - \frac{1}{2} U \Big[\rho^2 - \Big(\frac{\sin(\rho \pi)}{\pi } \Big)^2\Big]. 
\label{SFenergy}
\eeq
The factor of $1/2$ in the second term arises because the number of interaction terms are half the number of sites.

\subsection{Large-$U$ limit}
\begin{figure*}
\centering
\includegraphics[width=1.0\linewidth]{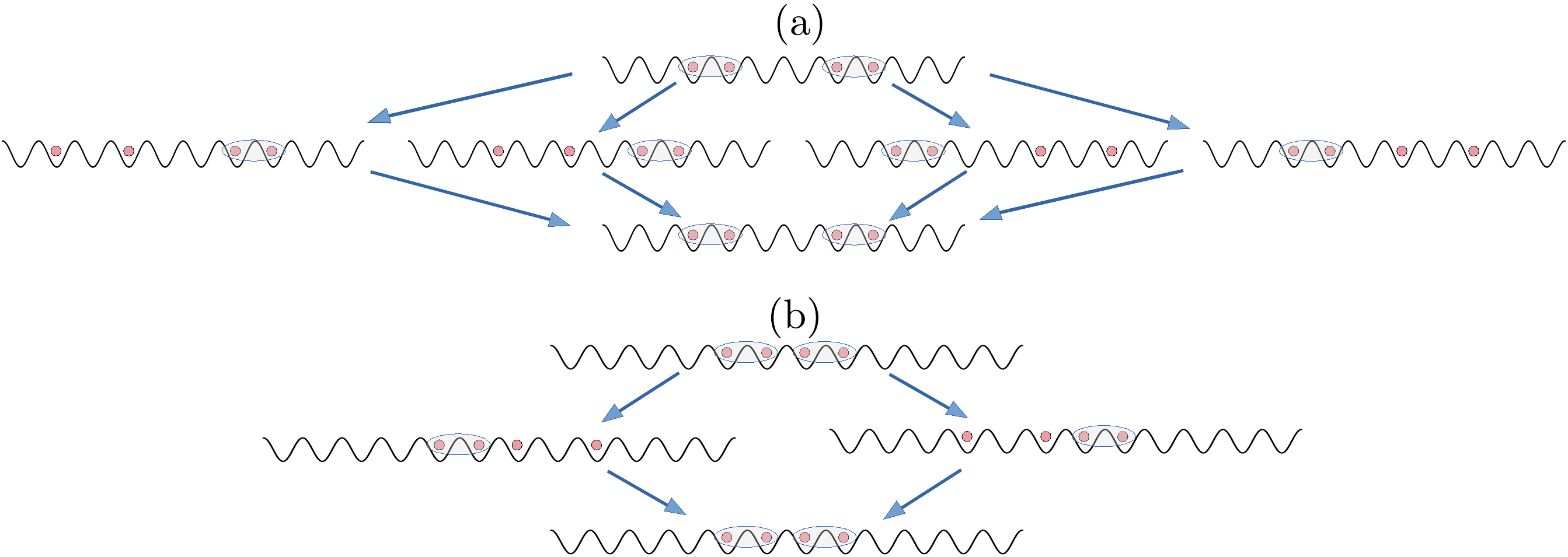}
\caption{(a) and (b) show schematic diagrams of the virtual hopping processes when two composite particles reside away from each other and on nearest-neighbor sites, respectively. These processes lead to an effective
repulsion between the dimers.}
\label{fig:sup_figure8}
\end{figure*}

We now work in the $U\rightarrow \infty$ limit where the dimers form a gapless state. The effective dimer Hamiltonian is
\beq
H_{\text{eff}} = \sum_p - \Big( \tilde{t} d^\dagger_{p} d_{p+1} + \text{H.c.} \Big) + V_{\text{eff}} n^d_{p} n^d_{p+1} - \mu_{\text{eff}} n_p, 
\label{eq:highUeq}
\eeq
where $\tilde{t} = \frac{2t^4}{U^3}, V_{\text{eff}} = \frac{2t^2}{U}$, and $\mu_{\text{eff}} = U + \frac{2t^2}{U}$.
In the next subsection we develop the analytic understanding of these hopping and interaction scales.

\subsection{Analytic understanding of interactions and hoppings of dimer}
\label{appA}

In this section, we will first calculate the energy of a dimer, i.e., a state in which two particles sit next to each other at sites $(j,j+1)$.
The energy of a dimer is
just equal to $-U$ if we ignore the hopping $t$. Then, considering the limit $U \gg t$, we will
calculate the lowest order correction to the energy
by using perturbation theory in terms of the small parameter $t/U$.
There will be a contribution from a process in which the
rightmost particle hops with an amplitude $-t$ from the site $j+1$ to $j+2$ and then back to $j+1$. The state
with particles at $(j,j+2)$ has energy equal to 
zero. Hence, to second order in perturbation theory, we get a correction
to the energy equal to $(-t)^2/(-U) = - t^2/U$. Similarly, there will
be a correction equal to $-t^2/U$ from a process in which the leftmost
particle hops from the site $j-1$ to $j$ and back. Adding these
two contributions, we see that the energy of a dimer is
\beq E_{\text{1 dimer}} ~=~ - ~U ~-~ \frac{2t^2}{U}. \label{emmer} \eeq
Eq.~\eqref{emmer} is the energy of an isolated dimer, i.e., when
there is no dimer to its immediate left or right. We now consider
what happens if there are two dimers sitting next to each other,
with particles at sites $(j,j+1,j+2, j+3)$. The zeroth-order energy of this state is $-2U$.
Now, to second order in $t$, there is a contribution from the rightmost
particle hopping from the site $j+3$ to $j+4$ and back, and, similarly,
from the leftmost particle hopping from $j$ to $j-1$ and back. Hence,
the energy of two dimers sitting next to each other is
\beq E_{\text{2 dimers}} ~=~ - ~2U ~-~ \frac{2t^2}{U}. \label{e2mmer} \eeq
Comparing this with twice the energy of a single dimer given in 
Eq.~\eqref{emmer}, we see that there is effectively a repulsive 
interaction equal to $V = 2t^2/U$ between two neighboring dimers. This implies that comparing these energies with the effective dimer Hamiltonian terms ($ V_{\text{eff}} n^d_{p} n^d_{p+1} - \mu_{\text{eff}} n_p$) (see Eq.~\eqref{eq:highUeq}) leads to
\beq
\mu_{\text{eff}} = U + \frac{2t^2}{U} ~~~{\rm and}~~~ V_{\text{eff}} = \frac{2t^2}{U}. \eeq
These virtual processes which lead to the effective repulsion  $V_{\text{eff}}$ between dimers are also illustrated in~Fig.~\ref{fig:sup_figure8}. 

Next, we will study the motion of an isolated dimer. Let us consider the effective hopping amplitude of
a dimer to move from the sites $(0,1)$ to $(2,3)$. It is clear that a
minimum of 4 hoppings is required to go between these two states.
Perturbation theory then suggests that the effective hopping 
amplitude of a dimer must be of order $t^4/U^3$ to leading
order in the ratio $t/U$. We can make this more
precise as follows. Let us label all the possible states
of two particles on four sites in terms of their positions $(n_1,n_2)$, where $0 \le n_1 < n_2 \le 3$. We find that there are 6 such states, which are denoted by $(0,1)$, $(0,2)$,
$(0,3)$, $(1,2)$, $(1,3)$ and $(2,3)$.
Certain pairs of these states are connected to each other by the
hopping amplitudes $-t$. For instance, the states $(0,1)$ and 
$(0,2)$ are connected to each other, as are $(0,2)$ and $(1,2)$.
This gives the off-diagonal matrix elements of a $6 \times 6$ Hamiltonian $H$.
Further, $H$ has two diagonal matrix elements equal to $-U$ for
the dimer states $(0,1)$ and $(2,3)$. Next, we numerically diagonalize
this Hamiltonian. For $U \gg t$, we indeed find that the lowest two states 
have energies very close to $-U$, and the gap between them, $\Delta E$, is of the order of $t^4/U^3$.
(We find that $\Delta E \simeq 4 ~t^4/U^3$ for $U \gg t$).
The corresponding eigenstates have wave functions which are peaked at
the states $(0,1)$ and $(2,3)$; the wave functions 
of the two states have the same sign in the lower energy state 
and opposite signs in the higher energy state.
We can now define an effective $2 \times 2$ Hamiltonian
for these two states which has off-diagonal entries equal to 
$\ga = \Delta E/2$ and diagonal entries equal to $-U$.
We thus conclude that a dimer has an effective hopping amplitude equal to 
$\tilde{t} \simeq 2 ~t^4/U^3$ (see Eq.~\eqref{eq:highUeq}).

\subsection{Comparison of small-$U$ and large-$U$ limits}
\begin{figure}[b]
\centering
\includegraphics[width=1.0\linewidth]{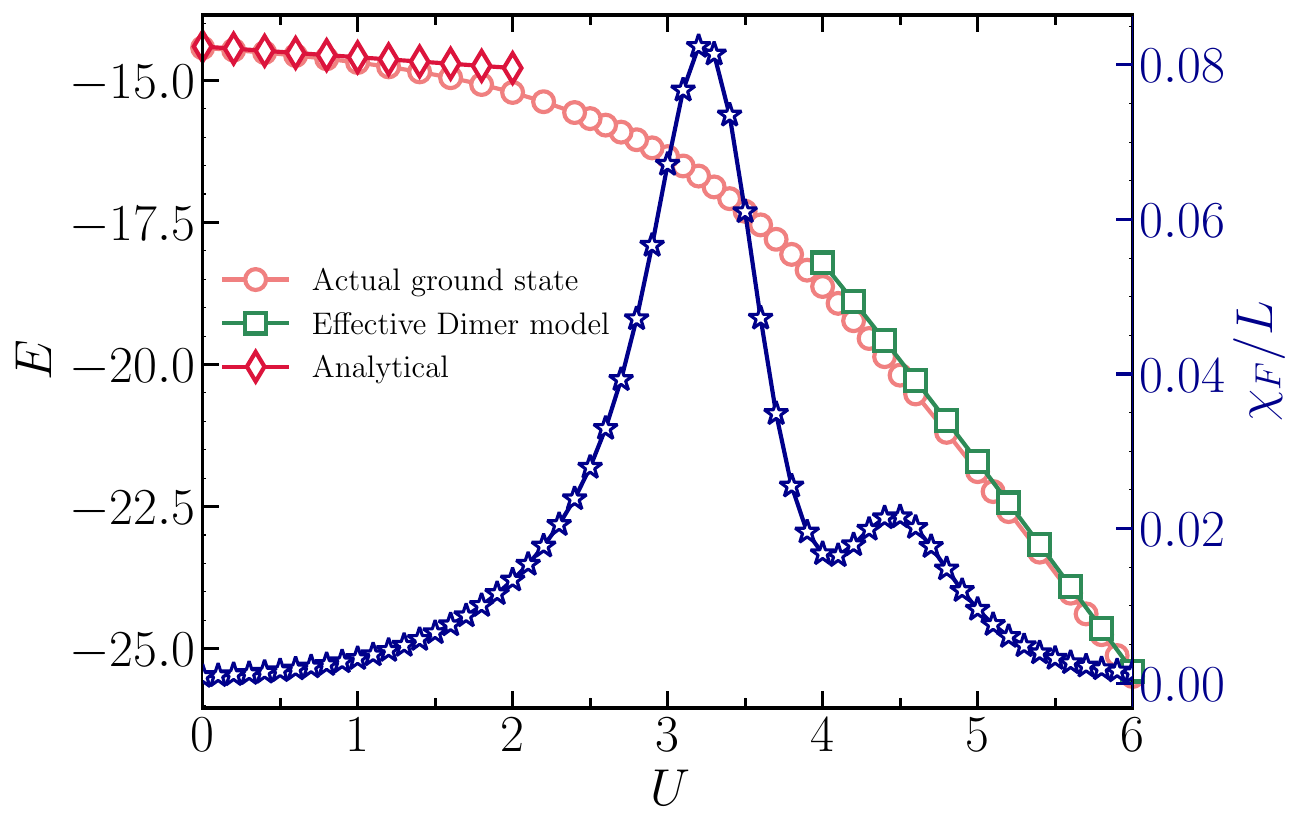}
\caption{Ground state energies obtained from three different approaches for a system of size $L=32$ at $1/4$ filling under PBC are shown on the left axis, namely, the low-$U$ Fermi-sea expression given in Eq.~\eqref{eq:lowUeq} (red diamonds), the energies from the effective dimer Hamiltonian in Eq.~\eqref{eq:highUeq} (green squares), and the results from the actual Hamiltonian in Eq.~\eqref{eq:Ham} (red circles). $\chi_F/L$ is plotted as a function of $U$ and displayed on the right axis using blue stars.}
\label{fig:sup_figure1}
\end{figure}

The ground state energy for a given filling can be compared with the estimates obtained from Eq.~\eqref{eq:lowUeq} and Eq.~\eqref{eq:largeUlim}. The comparison of the actual ground state energy with these two limits is shown in Fig.~\ref{fig:sup_figure1}. 
In order to decipher if there is a phase transition between these two states, we also study the fidelity susceptibility as given by
\beq
\chi_{F}(U) = \lim_{(U-U^{'})\to 0}\frac{-2\ln|\langle\Psi(U)|\Psi(U^{'})\rangle|}{(U-U^{'})^2}
\eeq
where $|\Psi(U^{'})\rangle$ is the ground state at any $U'$.
The two susceptibility peaks clearly show the formation of an intermediate phase.

\subsection{Additional Numerical Results}

\subsubsection{Estimation of $U_{c1}$}

\begin{figure}[t]
\centering
\includegraphics[width=1.\linewidth]{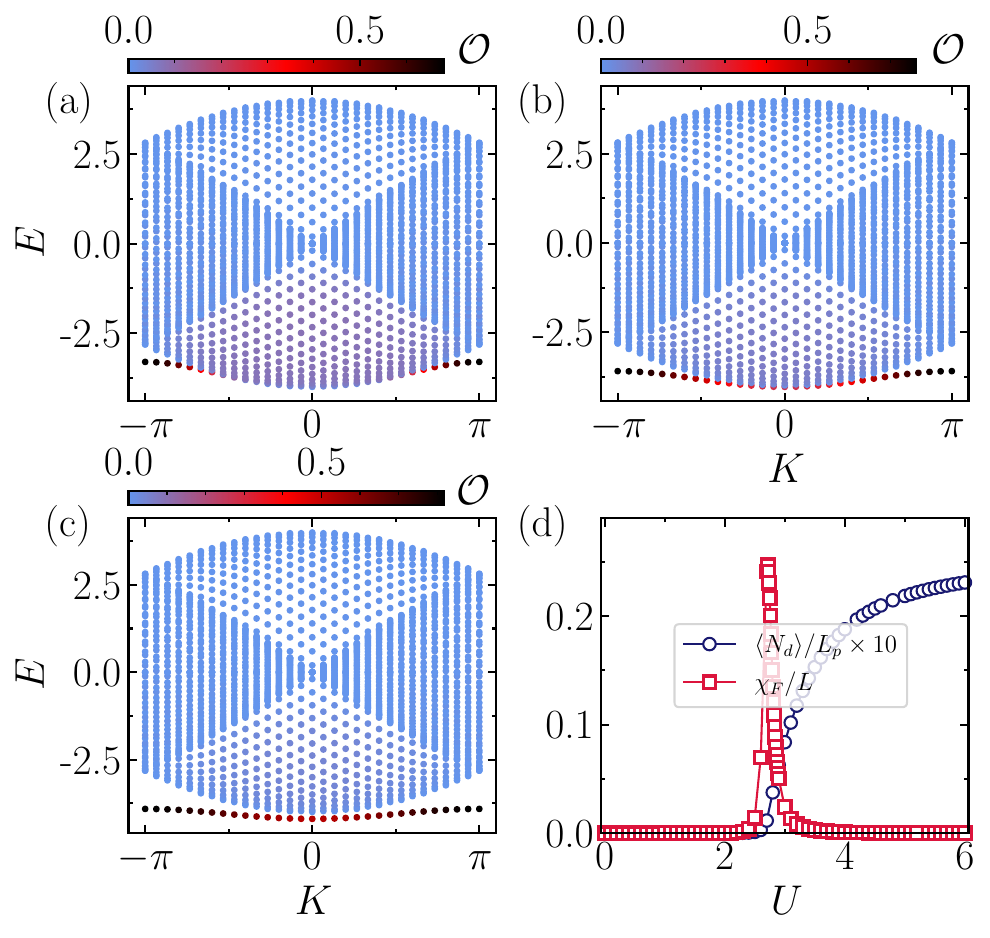}
\caption{(a), (b), and (c) show the band structures of two particles for $U=2.5,~2.9$ and $3.3$ respectively. The color indicates the overlap of the eigenstates with the bound states of two bosons residing on nearest-neighbor sites. (d) shows the values of $\chi_F/L$ (red squares) calculated under OBC and $\langle N_d\rangle/L_p$ (blue circles) calculated under PBC for two particles as a function of $U$ calculated using a system of $L=80$ sites. $\langle N_d\rangle/L_p$ has been magnified ten times for better visibility.}
\label{fig:sup_figure5}
\end{figure}

Comparing the energy of the atomic superfluid (see Eq.~\eqref{SFenergy}) with that of uncorrelated dimers with energy 
$ - (U + 2\frac{t^2}{U}) \frac{\rho}{2}$ gives $U_{c1} \sim 3.1$. To confirm this numerically, we plot the dimer density $\langle N_d\rangle/L_p$, taking two particles in a system of $L=80$ sites and show in Fig.~\ref{fig:sup_figure5}(d) using blue circles.
It is evident that the dimer density starts being finite at $U\sim2.5$.
This also agrees well with the peak in the fidelity susceptibility, which is shown using red squares.

However, this value is different than the predicted value of $U$ ($U\sim 3.1$) for the transition from atomic SF to DSF for a finite density of particles. To understand this we study the bound state formation in more detail in the next subsection. 

\subsubsection{Bound state formation}

 We plot the band structures for two particles for three different values of $U=2.5,~2.9,~\text{and},~3.3$, and show in Fig.~\ref{fig:sup_figure5}(a), (b), and (c), respectively.
The colorbar here indicates the overlap ($\mathcal{O}$) between the eigenstate and the two-particle bound state on nearest-neighbor even composite sites.
One can clearly see that for $U=2.5$, the finite value of $\mathcal{O}$ in the ground state indicates that the formation of a bound state has started.
With the increase in the value of $U$, the bound state band has now started getting separated from the scattering band, which can be seen from Fig.~\ref{fig:sup_figure5}(b) at $U=2.9$.
However, for $U=3.3$ [See Fig.~\ref{fig:sup_figure5}(c)], the bound state band is completely separated from the scattering band, which agrees well with the predicted value. 

\subsubsection{Estimation of $U_{c2}$}

Converting Eq.~\eqref{eq:highUeq} into fermions and calculating the ground state energy under Hartree-Fock approximation gives the energy
\bea
\epsilon_{\text{DSF}} &=& -~ \frac{4 t^4}{\pi U^3} ~\sin (\rho \pi) ~+~ \frac{2t^2}{U} ~\Big[\rho^2 - \Big(\frac{\sin(\rho \pi)}{\pi } \Big)^2\Big] \non \\
&& -~ \rho ~(U + \frac{2t^2}{U}), \label{lowUeq} \eea
A d-CDW state with uncorrelated dimers gives
\beq
\epsilon_{\text{d-CDW}} ~=~ - ~\rho ~( U + \frac{2t^2}{U} ).
\eeq
Comparing these two gives an estimate of $U_{c2} \sim 6.16$ for $\rho = 1/4$.

\begin{figure}[t]
\centering
\includegraphics[width=0.7\linewidth]{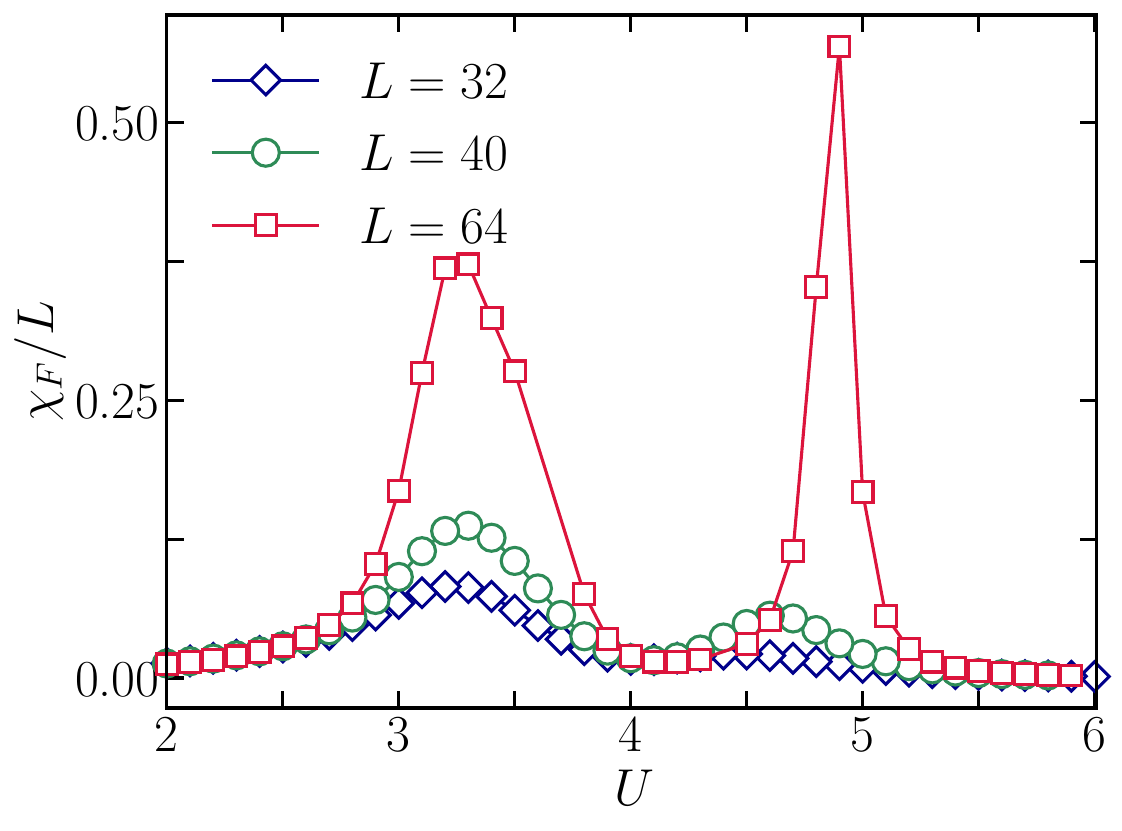}
\caption{Fidelity susceptibility($\chi_F/L$) plotted as a function of $U$ for various system sizes at $\rho=1/4$ filling under OBC.}
\label{fig:sup_figure3}
\end{figure}

\subsection{DMRG Details}

\subsubsection{Observables using DMRG}
Observables for larger system sizes are calculated using DMRG with OBC.
Hence, to avoid edge effects, the real space correlation values used for the calculation of $N_b(K)$, $N_d(K)$, and $S^d(K)$ are taken from $L/4$ to $3L/4$ sites in terms of original sites.
Accordingly, all the expressions described in the main text are changed as follows.
\beq
N_b(K) = \frac{2}{L} \Big(\sum_{i=L/4}^{3L/4-1} \sum_{j=L/4}^{3L/4-1} \langle b^\dagger_i b_{j} \rangle \exp(-i K r) \Big) \eeq
where $r=|i-j|$.
\beq
N_d(K) = \frac{2}{L_p} \Big(\sum_{p=L_p/4}^{3L_p/4-1} \sum_{q=L_p/4}^{3L_p/4-1} \langle d^\dagger_p d_{q} \rangle \exp(-i K r) \Big)
\eeq
where $r=|p-q|$ with $p$ and $q$ representing the composite sites.
\beq
S^d(k) = \frac{4}{L_p^2}\Big(\sum_{p=L_p/4}^{3L_p/4-1} \sum_{q=L_p/4}^{3L_p/4-1} 
 e^{i k r} \left(\langle n^d_p n^d_{q}\rangle
- \langle n^d_p \rangle \langle n^d_{q} \rangle\right).
\eeq

\subsubsection{System-size dependence}

In order to show that the two susceptibility peaks mentioned in the main text are thermodynamically stable, we plot the susceptibility values as a function of $U$ for different system sizes.
The increasing susceptibility peaks (see Fig.~\ref{fig:sup_figure3}) with increasing system size suggest that the two transitions are indeed thermodynamically stable.

\begin{figure}[]
\centering
\includegraphics[width=0.7\linewidth]{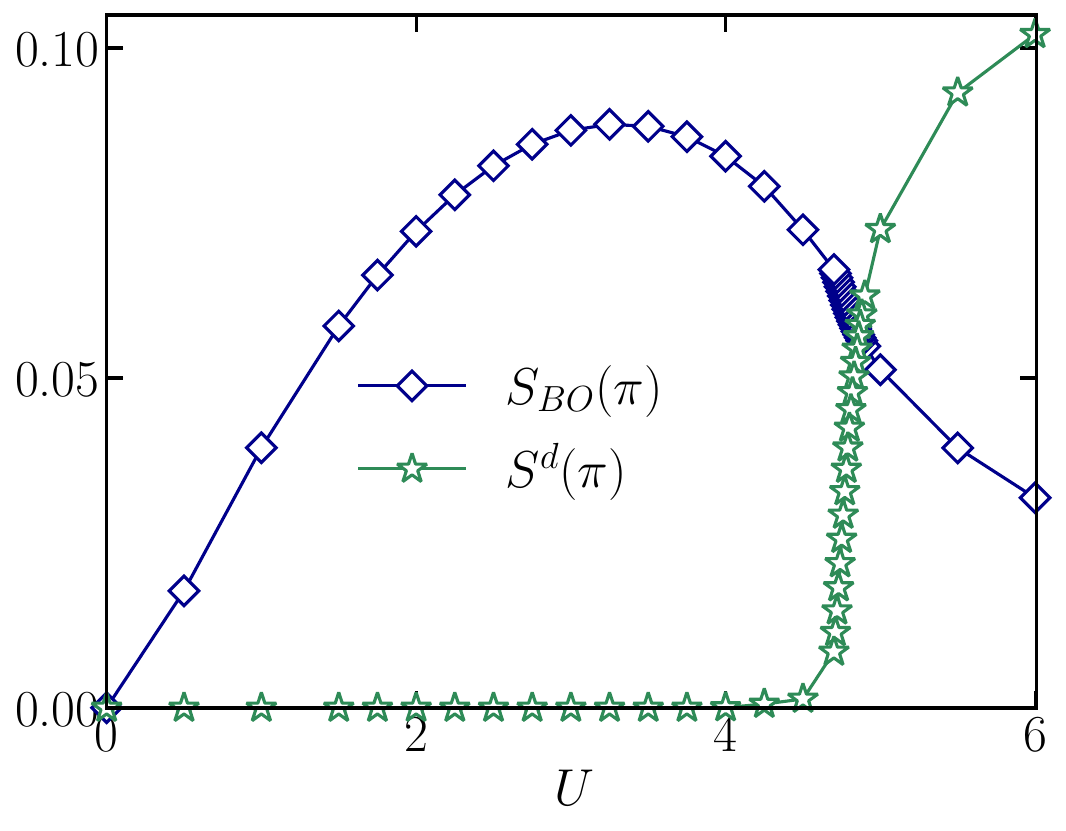}
\caption{The extrapolated values of the bond order structure factor ($S_{BO}(\pi)$) and dimer CDW structure factor ($S^d(\pi)$) has been plotted as a function of $U$ at half-filling under OBC.}
\label{fig:sup_figure2}
\end{figure}

\subsection{Physics at half-filling}

At half-filling, the system as a function of $U$ undergoes a single transition from a bond-ordered phase to a charge-density wave among the composite sites. This limit has been studied in Ref.~\cite{parida2025}; for completeness, we show that a single transition occurs at $U \sim 4.7$. The transition between bond-order and the dimer CDW can be captured within the corresponding structure factor as shown in Fig.~\ref{fig:sup_figure2}.

\begin{figure}[t]
\centering
\includegraphics[width=1.0\linewidth]{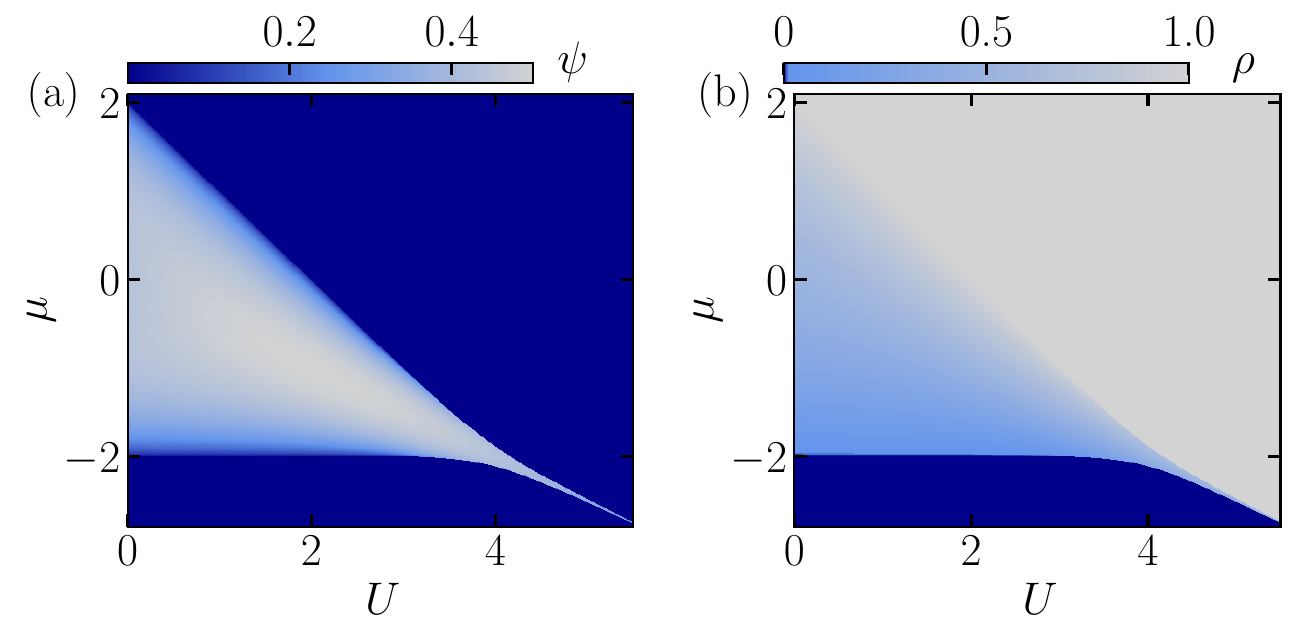}
\caption{The phase diagram in the $U-\mu$ plane calculated using the mean-field method by considering a cluster of two sites. The color scales in panels (a) and (b) represent the superfluid order parameter ($\psi=\langle b^\dagger\rangle$) and the particle density ($\rho=N/L$), respectively.}
\label{fig:sup_figure4}
\end{figure}

\begin{figure*}[t]
\centering
\includegraphics[width=1\linewidth]{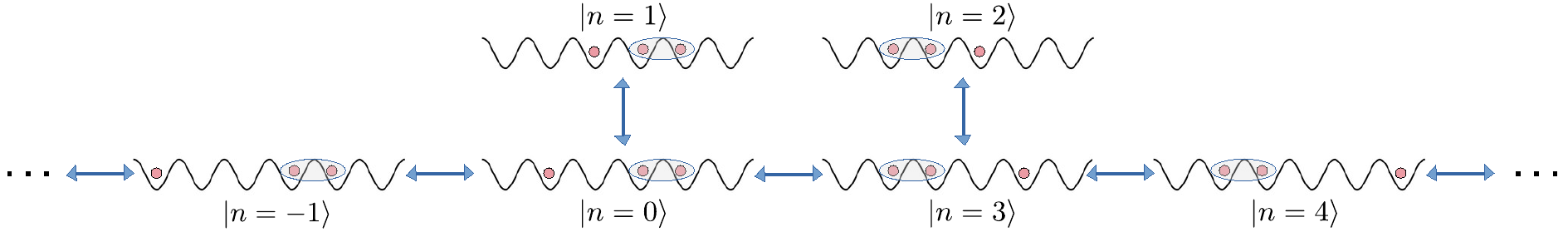}
\caption{Schematic diagram depicting the possible states, $| n \rangle$, of a dimer and a single particle (see Eq.~\eqref{boundstate}), and the connections between these
states such as from $|n=0\rangle$ to $|n=-1,~1,~3\rangle$ 
and from $|n=3\rangle$ to $|n=0,~2,~4\rangle$ (see Eq.~\eqref{connect}). Note that 
$|n=0\rangle$, $|n=3\rangle$ and their connections are related by the reflection symmetry $| n \ra \leftrightarrow |3 - n \ra$.}
\label{fig:sup_figure10}
\end{figure*}

\subsection{Mean-field results}

We perform a bosonic mean-field theory where we assume 
the
expectation values $\langle b^\dagger_i \rangle = \langle b_i \rangle = \psi$.

\begin{align}
 H_{\text{MF}} &= - ~t\sum_{i, i \in \text{even}} \Big( b_i^\dagger b_{i+1} +\text{H.c.} \Big) \notag \\ &~~~ - t \sum_{i, i \in \text{even}} \{ \psi (b_i + b^\dagger_i) + \psi (b_{i+1} + b^\dagger_{i+1}) \} \notag \\ &~~~ - U\sum_{i, i \in \text{even}} n_i n_{i+1} - \mu \sum_i n_{i}. \notag
 \label{eqn:Ham}
\end{align}

For every pair of sites the basis is given by $|00\rangle$, $|01\rangle$, $|10\rangle$, and $|11\rangle$. The MF Hamiltonian can be written in this basis as
\beq
\begin{array}{c||c|c|c|c||}
& |00\rangle & |01\rangle & |10\rangle & |11\rangle \\ \hline \hline
 |00\rangle & 0 & - t \psi & - t \psi & 0 \\ \hline
 |01\rangle & - t \psi & - \mu & -t & - t \psi \\ \hline
|10\rangle & - t \psi & -t & -\mu & - t \psi \\ \hline
|11\rangle & 0 & - t \psi & - t \psi & -U - 2 \mu \\
\hline
\end{array}
\eeq
A self-consistent solution allows us to stabilize the mean-field solution. While the mean-field captures the superfluid and the Mott insulating states with $\rho=0$ and $\rho=1$ as expected, none of the gapped states at half-filling, like the dimer-superfluid and phase-separated regions bond-ordered or the dimer CDW regions can be captured since these arise from virtual quantum processes. The results are shown in Fig.~\ref{fig:sup_figure4}.
The colorbars in Fig.~\ref{fig:sup_figure4}(a) and (b) represent the superfluid order parameter ($\psi$) and the particle density ($\rho$), respectively.
Comparing both the figures, it is evident that the superfluid order parameter is zero corresponding to the $\rho=0$ and $\rho=1$ regions, and hence clearly capturing the gapped phases at those densities.
However, between $\rho=0$ and $\rho=1$ regions, where the density is continuously varying, the corresponding superfluid order parameter is finite, and hence the gapped phases at half-filling are not captured within the mean-field approximation.

\subsection{Evidences of clustering}

\subsubsection{Three-particle bound state}

In this section, we will study the bound state of a dimer with a single 
particle. We will consider an infinitely long system with the Hamiltonian
\beq H ~=~ - t ~\sum_i ~(c_i^\da c_{i+1} ~+~ c_{i+1}^\da c_i) ~-~ U ~\sum_j~
n_{2j} n_{2j+1}. \label{ham1} \eeq
We will consider the limit $U \gg t$, and will consider configurations where
there is a dimer at a pair of sites $(2j,2j+1)$, and a single particle at a 
site $n$, where $n \ne 2j, 2j+1$. All such states have an energy $- ~U$ in the 
limit $t \to 0$. We will examine if there is a bound state in which the wave 
function goes to zero exponentially when the separation between the dimer and 
the particle goes to infinity.

Since the hopping of a dimer is of order $t^4/U^3$, we can take the
dimer to be stationary. We then find that it is sufficient to take the
dimer to be either at the sites $(0,1)$ or at ($2,3)$, and the set of states
can be labeled as $|n \ra$, such that

\begin{widetext}
 
\bea && {\rm either ~the ~dimer ~is ~at ~sites} ~(2,3)~ {\rm and ~the ~
particle ~is ~at} ~ n \le 1, \non \\
&& {\rm or ~the ~dimer ~is ~at ~sites} ~(0,1)~ {\rm and ~the ~particle 
~is ~at} ~ n \ge 2. \label{boundstate} \eea
We then find that the Hamiltonian connects these states with matrix element
$- ~t$ as follows:
\bea && |n\ra ~{\rm and} ~|n+1 \ra ~{\rm are ~connected~ if}~~ n \le -1 ~
{\rm or}~ n \ge 4, \non \\
&& |n=0\ra ~{\rm is ~connected ~to ~the ~three ~states} ~|n= -1, 1, 3\ra, 
\non \\
&& |n=3\ra ~{\rm is ~connected ~to ~the ~three ~states} ~|n= 0, 2, 4\ra.
\label{connect} \eea
\end{widetext}
These connections are shown by the schematic picture [Fig.~\ref{fig:sup_figure10}].

Note that the Hamiltonian has a reflection symmetry given by 
\beq | n \ra \leftrightarrow |3 - n \ra. \label{sym} \eeq
We now look for an eigenstate 
\beq | \psi \ra = \sum_{n=-\infty}^\infty c_n | n \ra \eeq
which satisfies $H | \psi \ra = E | \psi \ra$.
Given the matrix elements of $H$ in Eq.~\eqref{connect}, the energy eigenvalue equations are found to be
\bea -t ~(c_{n-1} ~+~ c_{n+1}) ~-~ U ~c_n &=& E ~c_n \non \\{\rm for} ~~n \le -1
~~{\rm and} ~~n \ge 4, \non \\
-t ~(c_{-1} ~+~ c_1 ~+~ c_3) ~-~ U~ c_0 &=& E ~c_0, \non \\
-t ~c_0 ~-~ U~ c_1 &=& E ~c_1, \non \\
-t ~c_3 ~-~ U~ c_2 &=& E ~c_2, \non \\
-t ~(c_0 ~+~ c_2 ~+~ c_4) ~-~ U~ c_3 &=& E ~c_3. \label{eigen} \eea
\begin{figure}[t]
\centering
\includegraphics[width=1.0\linewidth]{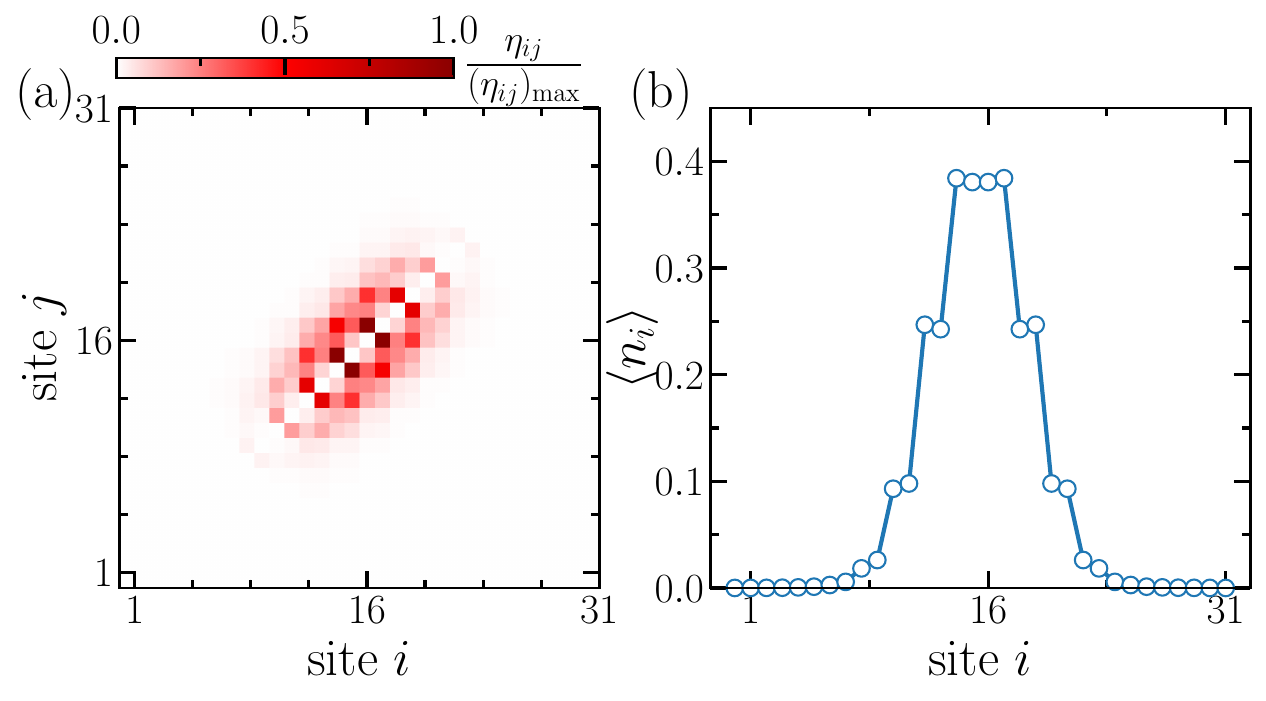}
\caption{(a) and (b) show the density-density correlations $\eta_{ij}/(\eta_{ij})_{\text{max}}$ and the on-site density $\langle n_i\rangle$, respectively, for three particles. We have taken $U = 30$ and $L=32$.}
\label{fig:sup_figure9}
\end{figure}
\begin{figure}[b]
\centering
\includegraphics[width=0.75\linewidth]{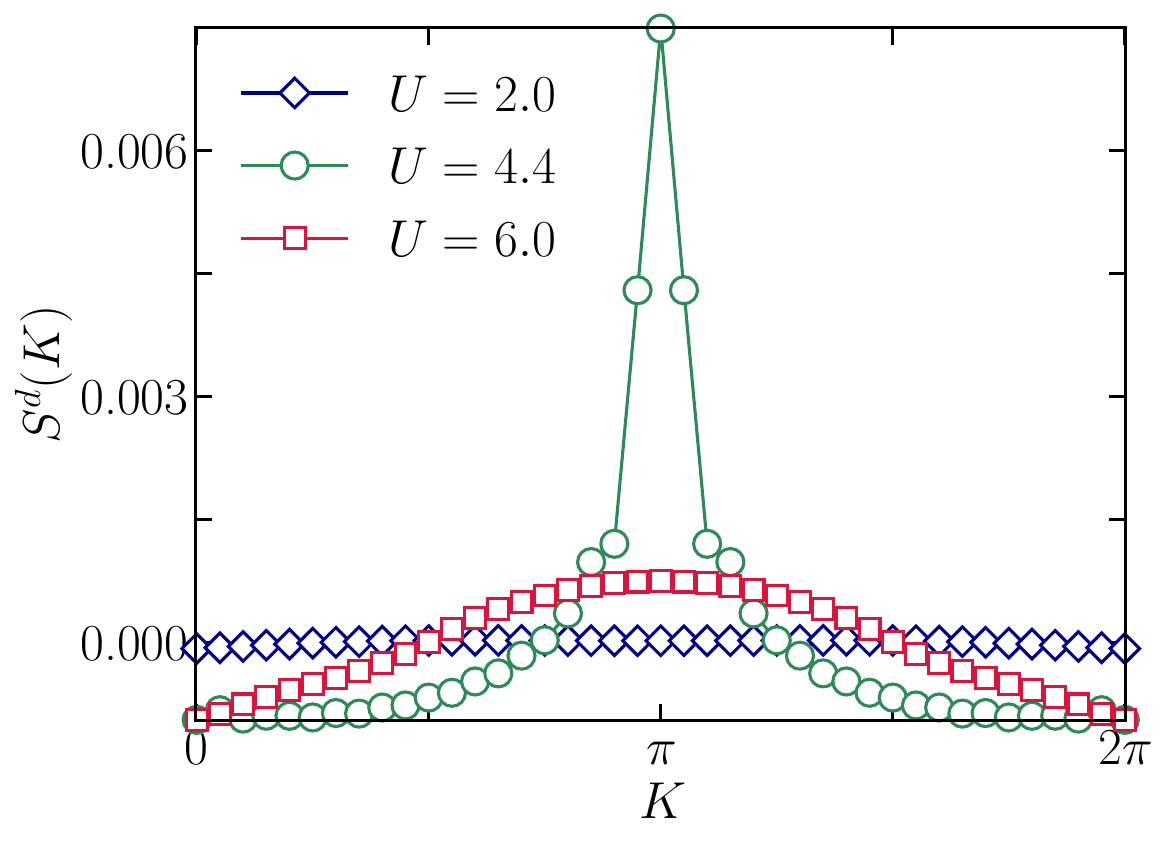}
\caption{The dimer CDW structure factor
$S^d(K)$ in the SF ($U=2.0$, blue diamonds), PS ($U=4.4$, green circles), and DSF ($U=6.0$, red squares) phases for $20$ particles in a system of $L=80$ sites.}
\label{fig:sup_figure6}
\end{figure}
Given the symmetry in Eq.~\eqref{sym}, we try a solution of Eq.~\eqref{eigen} 
of the form
\bea c_0 &=& c_3 ~=~ 1, \non \\
c_1 &=& c_2 ~=~ \beta, \non \\
{\rm and}~~ c_n &=& \al^{-n} ~~{\rm for}~~ n \le 0, \non \\
&=& \al^{n-3} ~~{\rm for}~~ n \ge 3, \label{guess} \eea
where we need $|\al| < 1$ to obtain a bound state.
Then the equations in Eq.~\eqref{eigen} for $n \le -1$ and $n \ge 4$ imply that the energy is given by
\beq E ~=~ -t ~(\al ~+~ \frac{1}{\al}) ~-~ U. \label{energy} \eeq
The equations for $c_0$ and $c_3$ in Eq.~\eqref{eigen} give
\beq E ~=~ -t ~(\al ~+~ \beta ~+~ 1) ~-~ U, \eeq
while the equations for $c_1$ and $c_2$ give
\beq E ~\beta ~=~ -t ~-~ U ~\beta. \eeq
Combining these equations, we obtain
\beq \beta ~=~ \frac{1}{\al} ~-~ 1, \eeq
and
\beq \al^3 ~+~ \al ~=~ 1. \label{alpha1} \eeq
Solving Eq.~\eqref{alpha1}, we find 
\beq \al ~\simeq~ 0.6823, \label{alpha2} \eeq
which implies that the bound state energy is
\beq E ~\simeq~ - ~U ~-~ 2.148 ~t. \eeq
This is slightly below the continuum of energies of the dimer and single
particle which is $- ~U~ - ~2 t$, hence it describes a stable bound state of a dimer and a single particle.

In order to show numerically that there is a three-particle bound state, we calculate the correlation matrix ($\eta_{ij}$) as well as the on-site particle density for three particles at $U=30$, shown in Figs.~\ref{fig:sup_figure9}(a) and (b) respectively.
The finite and large diagonal correlation near the central region indicates the formation of a bound state.
Consistently, the on-site probability density also shows a finite density near the center and exponential decay towards the boundaries.
\begin{figure}[t]
\centering
\includegraphics[width=1.0\linewidth]{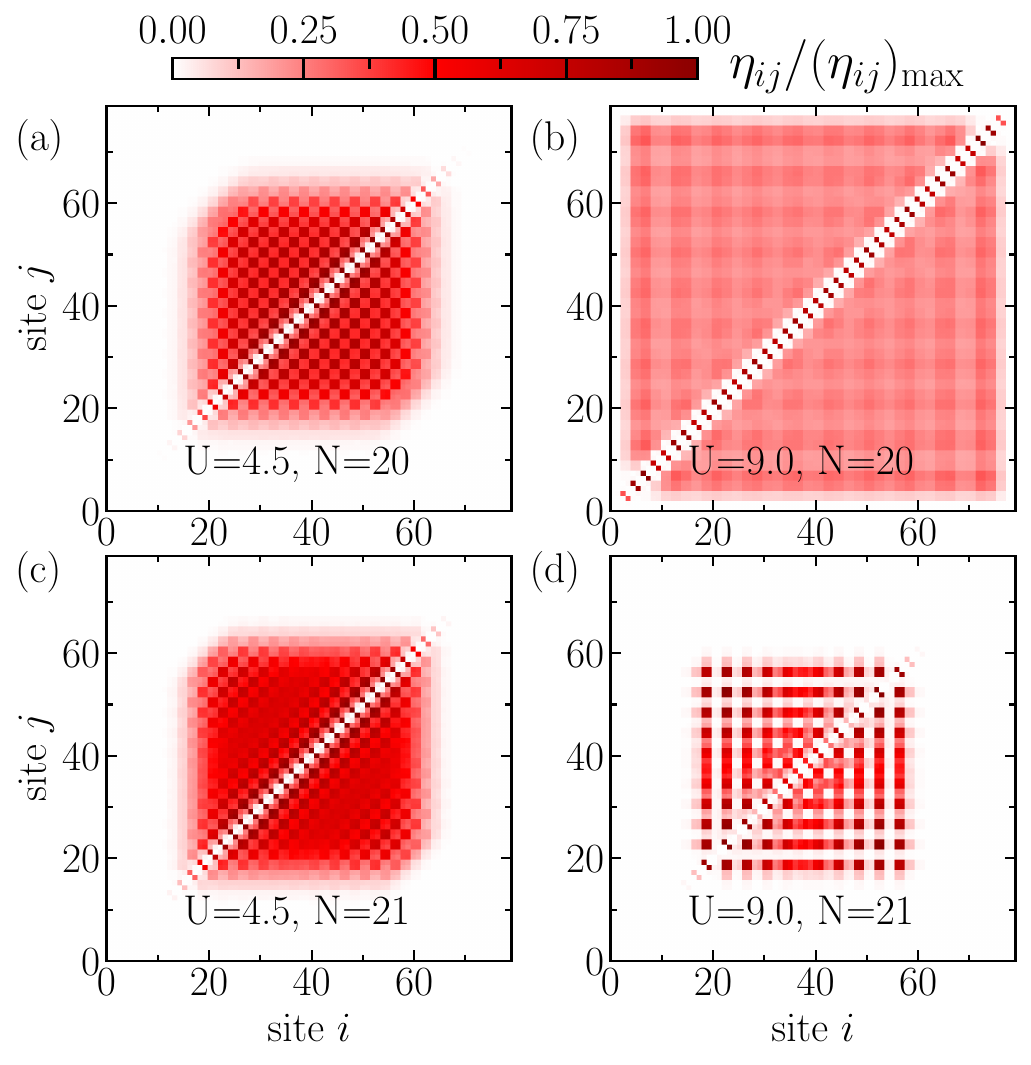}
\caption{(a) and (b) show the density-density correlations $\eta_{ij}/(\eta_{ij})_{\text{max}}$ when the particle number $N=20$ (even particle sector) for interaction strength $U = 4.5$ and $9.0$, respectively. (c) and (d) show the $\eta_{ij}/(\eta_{ij})_{\text{max}}$ with an odd number of particles, $N=21$, for interaction strength $U = 4.5$ and $9.0$, respectively. For all the calculations, we have considered a system size of $L=80$.}
\label{fig:sup_figure7}
\end{figure} 

\subsubsection{Local d-CDW order}

We would like to emphasize here
that away from the extremely dilute limit, when the
number of particles is odd, the system not only exhibits clustering (consistent with the three-particle calculation discussed above), but also develops a local d-CDW order starting from intermediate values of $U$ 
all the way up to the limit $U\rightarrow\infty$.
However, when the number of particles is even, this local d-CDW feature appears only for intermediate values of $U$.
This behavior is consistent with the features observed in the d-CDW structure factor $S^d(K)$, as discussed in the main text.
To illustrate the behavior of $S^d(K)$ across different regimes of the interaction strength, we consider systems with an even number of particles and plot $S^d(K)$ as a function of $K$ for three representative values of $U$, as shown in Fig.~\ref{fig:sup_figure6}. 
The $U$ values considered correspond to the SF phase ($U=2.0$), the intermediate PS region ($U=4.4$), and the DSF phase ($U=6.0$).
It is evident that a pronounced peak in $S^d(K)$ at $k = \pi$ appears only in the intermediate regime of $U$ ($U=4.4$). In contrast, no such peak is observed for the other interaction strengths, indicating the absence of d-CDW order in those cases.

In order to complement this observation, we present the correlation matrix ($\eta_{ij}$) for both even and odd numbers of particles in the system for $U=4.5$ 
(Figs.~\ref{fig:sup_figure7}(a) and (c)).
The correlation values are finite only in the central region and vanish elsewhere; this confirms the clustering in the system.
Moreover, the checkerboard pattern in the matrix 
$\eta_{ij}$ confirms the d-CDW ordering present in the system.
At higher values of $U$, the system
is a superfluid of dimers if it 
has an even number of particles. 
However, it turns out that the presence of an extra particle in the system always favors clustering with a local CDW puddle.
In order to demonstrate this, we plot $\eta_{ij}$ for both even and odd numbers of particles in the system at a larger value of $U=9.0$, as shown in Figs.~\ref{fig:sup_figure7}(b) and (d).
For the even particle sector 
(Fig.~\ref{fig:sup_figure7}(b)), the larger diagonal correlation indicates the presence of dimers, whereas the finite correlations across the entire lattice, except near the edges (a consequence of the OBC), confirms the superfluid nature of the system in this parameter regime.
In contrast, if the system has an odd number of particles, then even at larger values of $U$, there is always a lone particle present, and hence a local CDW pattern is always maintained as shown in Fig.~\ref{fig:sup_figure7}(d).

\subsection{Experimental realization} 

The model is equivalent to a spin-$1/2$ XXZ chain where the ZZ interaction appears only on the even bonds,
\begin{equation}
    \hat{H} = 
    -t \sum_{i=0}^{L-1} (\hat{S}^+_i \hat{S}^-_{i+1} + \text{h.c.}) 
    -U \!\sum_{i \text{ even}} (\hat{S}^z_i + 1/2) (\hat{S}^z_{i+1} + 1/2) \, .
    \label{eq:spin_model}
\end{equation}
Following \cite{Kuklov2003}, one can realize such a spin model by trapping two-component bosonic atoms in a spin-dependent optical lattice with strong onsite interactions. In the uniform case, the atoms are described by the Hamiltonian
\begin{align}
    \nonumber \hat{H} =&\; 
    - \sum_{i, \sigma} t_{\sigma} (\hat{a}_{i,\sigma}^{\dagger} \hat{a}_{i+1,\sigma} + \text{h.c.}) 
    - \sum_{i, \sigma} \mu_{\sigma} \hat{n}_{i, \sigma} \\
    & + \frac{1}{2} \sum_{i, \sigma, \sigma^{\prime}} U_{\sigma, \sigma^{\prime}} :\hat{n}_{a, \sigma} \hat{n}_{a, \sigma^{\prime}} : \, ,
\end{align}
where $\hat{a}_{i,\sigma}$ annihilates a boson of spin $\sigma \in \{ \uparrow, \downarrow\}$ at site $i$, $\hat{n}_{i,\sigma} \coloneqq \smash{\hat{a}_{i,\sigma}^{\dagger}} \hat{a}_{i,\sigma}$, and $:(\dots):$ denotes normal ordering. As shown in \cite{Kuklov2003}, for strong interactions at unit filling, the low-energy physics arises from virtual hoppings between neighboring sites, giving rise to the spin model
\begin{equation}
    \hat{H} = 
    - \sum_i \big[ J_{\perp} (\hat{S}^+_i \hat{S}^-_{i+1} + \text{h.c.}) + J_z \hat{S}^z_i \hat{S}^z_{i+1} \big] 
    - \sum_i B \hat{S}^z_i \, ,
\end{equation}
where 
\begin{align}
    & J_{\perp} = \frac{t_{\uparrow} t_{\downarrow}}{U_{\uparrow\downarrow}} \, , \quad
    J_z = \sum_{\sigma} t_{\sigma}^2 \left( \frac{2}{U_{\sigma\sigma}} - \frac{1}{U_{\uparrow\downarrow}} \right) \, , 
    \label{eq:Jz} \\[0.2em]
    & B = \mu_{\uparrow} - \mu_{\downarrow} + 4 \big( t_{\uparrow}^2 / U_{\uparrow\uparrow} - t_{\downarrow}^2 / U_{\downarrow\downarrow} \big) \, .
    \label{eq:B}
\end{align}
This approach has been implemented experimentally to realize XXZ chains with tunable anisotropies \cite{Jepsen2020}. To obtain the staggered model in Eq.~\eqref{eq:spin_model}, $J_z$ has to vanish on all the odd bonds. Firstly, this requires the coefficients of $\smash{t_{\uparrow}^2}$ and $t_{\downarrow}^2$ in Eq.~\eqref{eq:Jz} to have opposite signs. For $U_{\uparrow\downarrow} > 0$, we need $0 < U_{\uparrow\uparrow} < 2 U_{\uparrow\downarrow}$ and $U_{\downarrow\downarrow} < 0$ or $U_{\downarrow\downarrow} > 2 U_{\uparrow\downarrow}$. One can operate in this regime by tuning the inter- and intra-species scattering lengths via Feshbach resonances (see \cite{AmatoGrill2019} for ${}^7$Li atoms). To make $J_z = 0$, the hopping amplitudes $t_{\sigma}$ can be adjusted by varying the potential depths of the optical lattice \cite{Zwerger2003}. Second, for the even-odd modulation one can use a superlattice potential \cite{Yang2017} where the barrier height across the even bonds is slightly lower for spin-$\uparrow$ atoms (to increase $t_{\uparrow}$) and slightly higher for spin-$\downarrow$ atoms (to decrease $t_{\downarrow}$), such that $J_z$ becomes positive without changing $J_{\perp}$. One can also adjust the onsite potentials $\mu_{\sigma}$ \cite{Zupancic2016} to set $B=J_z/2$ on all bonds, as in Eq.~\eqref{eq:spin_model}. The effective interaction $U/t$ can be varied over a wide range using the sensitive dependence of $t_{\sigma}$ on the potential depth \cite{Zwerger2003, Bloch2005} and the large tunability of $U_{\sigma, \sigma^{\prime}}$ provided by Feshbach resonances \cite{AmatoGrill2019, Chin2010}.


\end{document}